\documentclass[a4paper,UKenglish,cleveref, autoref]{lipics-v2019}

\graphicspath{{./.}}

\usepackage[normalem]{ulem}
\usepackage{multirow}
\usepackage{booktabs} 
\usepackage{xspace}
\usepackage{graphicx}
\usepackage[colorinlistoftodos]{todonotes}
\usepackage{siunitx}

\newcommand{\cvf}{\ensuremath{cvf}\xspace}

\newcommand{\br}[1]{\ensuremath{\langle #1\rangle}\xspace}
\newcommand{\actnode}{active-node\xspace}
\newcommand{\pasnode}{passive-node\xspace}

\title{Benefits of Stabilization versus Rollback in Eventually Consistent Key-Value Stores}

\titlerunning{Stabilization vs Rollback for Key Value Stores}

\author{Duong Nguyen}{Michigan State University, USA}{nguye476@cse.msu.edu}{https://orcid.org/0000-0003-4894-5217}{}

\author{Sandeep S. Kulkarni}{Michigan State University, USA}{sandeep@cse.msu.edu}{}{}

\authorrunning{D. Nguyen and S.\,S. Kulkarni}



\category{}

\relatedversion{}

\supplement{The source code and experimental results are available at \url{https://sourceforge.net/projects/opodis2019duongnguyen/}.}

\acknowledgements{We want to thank the late Professor Ajoy K. Datta for the idea used in this paper.}

\nolinenumbers 

\EventEditors{John Q. Open and Joan R. Access}
\EventNoEds{2}
\EventLongTitle{42nd Conference on Very Important Topics (CVIT 2016)}
\EventShortTitle{CVIT 2016}
\EventAcronym{CVIT}
\EventYear{2016}
\EventDate{December 24--27, 2016}
\EventLocation{Little Whinging, United Kingdom}
\EventLogo{}
\SeriesVolume{42}
\ArticleNo{23}

\begin{document}

\maketitle

\begin{abstract}
In this paper, we evaluate and compare the performance of two approaches, namely self-stabilization and rollback, to handling consistency violation faults (\cvf) that occurred when a distributed program is executed on eventually consistent key-value store.
We observe that self-stabilization is usually better than rollbacks in our experiments. Moreover, when we aggressively allow more \cvf in exchange of eliminating mechanisms for guaranteeing atomicity requirements of actions, we observe the programs in our case studies achieve a speedup between 2--15 times compared with the standard implementation.
We also analyze different factors that contribute to the results.
Our results and analysis are useful in helping a system designer choose proper design options for their program.

\end{abstract}

\section{Introduction}
A traditional distributed system consists of a set of \textit{nodes} connected to each other by a set of channels. Each node is associated with a set of \textit{variables} and a set of \textit{actions}. These actions read the variables (of that node and possibly other nodes) and update the variables. Continued execution in this manner updates the variables of (possibly) all nodes in a manner as desired by the goals of the system.
For example, consider a program for distributed maximal matching. In such a program, each node is associated with its matching partner (if any) and other variables to keep track of proposals, timestamps, etc. The program reads these variables to identify which pairs of nodes should be matched with each other. We denote this model as an \textit{active node model}. Intuitively, the reason is that the nodes appear to be active in updating their own state. 

In \cite{NKD2019ICDCN}, we introduced the notion of a \textit{passive node model} that is targeted towards scenarios where the number of nodes is very large. For example, if we wanted to perform the matching algorithm in a graph of tens or hundreds of thousands of nodes, clearly, having such large active nodes is not feasible. In the passive node model, the variables associated with the nodes are stored in some data store, which in turn could be replicated and/or partitioned. A set of clients operate on this data store to perform the actions as described by the program. For example, in a passive node model for the matching program, the variables associated with the nodes would be stored in a data store. The \textit{clients} will read the relevant variables to determine if matching of one or more node should be changed. The process will continue until a maximal matching is found. 

In the active state model, it is assumed that when a node reads the variables of its neighbors, it obtains the latest information about that node. This is reasonable given that there is only one copy of each variable. And, this copy is stored with the node that owns that variable. The same property could be achieved if we use a single copy of the data store in the passive node model. However, maintaining multiple copies is beneficial for many reasons such as fault-tolerance, improved access time and availability of data. When multiple copies are maintained, if we provide sequential consistency, then this datastore appears as a single copy. Hence, in this case, each client will obtain the latest copy of the data. However, sequential consistency requires high overhead whereas a weaker consistency model, e.g., eventual consistency, can substantially increase throughput and reduce latency. However, if we use eventual consistency then a client may obtain stale information about the status of nodes. This is denoted as consistency violation fault (\cvf). Preventing \cvf{s}, i.e., preventing access to such stale information essentially requires the use of sequential consistency thereby resulting in an increase of access time and reduction in throughput. 

Since preventing the use of stale information is not desirable, there are two ways of dealing with such stale information. One approach is to detect the use of such stale information. This can be achieved by using algorithms such as those in \cite{CG98DC}. To achieve this, we need to run a monitor concurrently with the program. When the monitor detects a violation, we can restore the program to a previous state and continue the execution thereafter. In \cite{NCKD2018LADC}, we found that the detection of the violation is often very quick. Specifically, it is possible to restart a few actions taken by clients involved in conflicting access to data rather than requiring all clients to rollback in a coordinated fashion. (We refer the reader to Section \ref{sec:detect-rollback} for details.)
An alternate approach is to use stabilization \cite{EDW426}. A stabilizing program is guaranteed to recover from an arbitrary state to a legitimate state. Thus, if a client ends up updating the information of some node based on stale information, we can treat it as if a fault caused the state of that node to be perturbed. A stabilizing program is designed to recover from such a fault as long as such faults do not occur frequently. In particular, a stabilizing program is guaranteed to recover after the faults stop (or if faults for a long enough time). If the \cvf{s} occur frequently then the program may not be able to recover from them. At the same time, given the nature of \cvf{s}, expecting them to never occur during recovery is not reasonable. Thus, we need to evaluate how \cvf{s} perturb the recovery to determine the overall effect. In \cite{NKD2019ICDCN}, it is shown that the perturbation caused by \cvf{s} is not as severe in that tolerating \cvf{s} and using eventual consistency is better than eliminating \cvf{s} with sequential consistency. 

\textbf{Summary of the main results.} 
In this paper, we focus on the tradeoff between these two approaches. Clearly, if the underlying program is not stabilizing then we must utilize the rollback-recovery based approach. Hence, we consider stabilizing programs where both approaches are feasible. We run several distributed graph-based applications on LinkedIn's Voldemort key-value store on our local network and Amazon AWS network.
From the analysis of experimental results, we observe the followings:
\begin{itemize}
    \item For the case study applications used in these experiments, namely planar graph coloring, arbitrary graph coloring, and maximal matching, we observe that the stabilization approach is better than the detect-rollback approach, especially when we treat violations of atomicity requirement (e.g. violation of mutual exclusion where the same data item could be concurrently read and updated by two clients) as \cvf{s}. Specifically, the stabilization approach improves the convergence time (compared to running on sequential consistency) by \SI{25}{\percent} to \SI{35}{\percent}. With aggressive stabilization approach, the speedup in convergence time is between 2 to 15 times. By contrast, the detect-rollback approach improves convergence time by 30\% in the best case, and potentially causes performance to suffer when compared with sequential consistency.
    The main reason for this difference is that securing the atomicity of actions contribute a significant amount of time in the computation. By eliminating this requirement, the computation time is improved. On the other hand, without the atomicity requirement, the clients can read inconsistent data. In the self-stabilization approach, these inconsistencies can be treated as \cvf{s} which result in additional recovery time. However, this overhead is outweighed by the computation time reduction due to the elimination of the atomicity requirement. By contrast, since the rollback recovery approach requires these mechanisms to detect possible violations and trigger the rollback, this approach is not able to utilize such benefit of eliminating atomicity requirement.
    \item We analyze different factors that affect the performance of both approaches such as type of graph input and type of applications. We find that on graphs with complex connectivity between nodes such as social graphs, the overhead for providing atomicity requirement such as mutual exclusion is high, and the stabilization approach performs well. By contrast, the detect-rollback approach suffers from higher chance of violations as well as a significant amount of work wasted during rollbacks. In regular graphs, the overhead of mutual exclusion is reduced and the detect-rollback provides some benefit.
    \item Although more beneficial, the stabilization approach could suffer from some rare \cvf{s} that prevent the application to converge. In such cases, we propose some heuristics to improve the performance of stabilization such as randomization and tracking states of active neighbor nodes.
\end{itemize}

\textbf{Contributions of the paper.} To the best of our knowledge, our paper is the first to analyze the trade-off between the two approaches of handling data anomalies when running graph-based applications on eventually consistent data stores. We find that when the self-stabilization option is available, it usually provides a better benefit than the detect-rollback recovery approach. Moreover, if we aggressively disable mechanisms for atomicity requirements of actions and treated the violations as other \cvf faults, the stabilization approach outperforms the detect-rollback approach. However, in some applications, self-stabilization algorithms do not exist. In such circumstances, the rollback approach may be the choice. We also analyze different factors that influence the performance of each approach such as the type of application, the characteristic of the input. Our analysis may be useful for systems designers who have to consider different design options for their programs on distributed key-value stores.

\textbf{Organization of the paper.} In Section \ref{sec:sys-model-architecture}, we present the system models/architecture, the definition of \cvf, and briefly recall the stabilization and detect-rollback approach. In Section \ref{sec:expriment}, we present and analyze the experimental results. Finally, we conclude the paper in Section \ref{sec:concl}.

\section{System Model/Architecture}
\label{sec:sys-model-architecture}
In this section, we recall some important notions used in this paper that have been introduced in \cite{NKD2019ICDCN, NCKD2018LADC}. Specifically, Section \ref{sec:distprogram} defines the notion of distributed programs and discuss how 
the computations of these programs are represented in the traditional \actnode model and the \pasnode model.
Section \ref{sec:stabilization} recalls the definition of the notion stabilization. In Section \ref{sec:voldemort-kv}, we describe the architecture of Voldemort key-value store and how it implements the \pasnode model. Next, we describe consistency violating faults (\cvf) which are caused by data anomalies in eventual consistency. Finally, we discuss the detect-rollback approach for handling \cvf in Section \ref{sec:detect-rollback}.

\subsection{Distributed Programs: Active and Passive Node Model}
\label{sec:distprogram}
A program $p$ consists of a set of nodes $V_p$ and a set of edges $E_p$. We assume that $\forall i \in V_p, (i,i) \in E_p$. Each node, say $j$, in $V_p$ is associated with a set of variables $var_j$. The union of all node variables is the set of variables of the program $p$, denoted by $var_p$.
A state of $p$ is obtained by assigning each variable in $var_p$ a value from its domain. 
State space of $p$, denoted by $S_p$, is the set of all possible states of $p$. 

Each node $j$ in program $p$ is also associated with a set of actions $ac_j$. An action in $ac_j$ is of the form $g \longrightarrow st$, where the guard $g$ is a predicate involving $\{ var_k :  (j, k) \in E_p \}$ and $st$ updates one or more variables in $var_j$.
We say that an action $ac$ (of the form $g \longrightarrow st$) is enabled in state $s$ if and only iff $g$ evaluates to true in state $s$. A node $j$ is said to be enabled at state $st$ if any action in $ac_j$ is enabled in $st$.
The transitions of action $ac$ (of the form $g \longrightarrow st$) are given by  $\{ (s_0, s_1) |$  $s_0, s_1 \in S_p$, $g$ is true in $s_0$ and $s_1$ is obtained by execution $st$ in state $s_0$\}. 
Transitions of node $j$ (respectively, program $p$) is the union of the transitions of its actions (respectively, its nodes). We use $\delta_{ac}, \delta_j$ and $\delta_p$ to denote transitions corresponding to action $ac$, node $j$ and program $p$ respectively. 
%


\textbf{Computation in traditional/active node model. }
In the traditional/\actnode model, the computation program $p$ is of the form $\br{s_0, s_1, \cdots}$ where 

\begin{itemize}
\item  $\forall l: l \geq 0: $, $s_l$ is a state of $p$,
\item $\forall l: l \geq 0: (s_l, s_{l+1})$ is a transition of $p$ or\\
($(s_l = s_{l+1})$ and no action of $p$ is enabled in state $s_l$), and
\item If some action $ac$ of $p$ (of the form $g \longrightarrow st$) is continuously enabled (i.e., there exists $l$ such that $g$ is true in every state in the sequence after $s_l$) then $ac$ is eventually executed (i.e., for some $x \geq l$, $(s_x, s_{x+1})$ corresponds to execution of $st$.) 

\end{itemize}

We note that the above computation model corresponds to centralized daemon with interleaving semantics. It can be extended to other models and semantics as well \cite{NCKD2018LADC,NKD2019ICDCN}.

The resulting computation guarantees that two neighboring nodes do not execute simultaneously. In turn, the resulting computation is realizable in the original model.
(Our observations/results are also applicable to other models such as powerset semantics, distributed daemon, etc.)


As discussed in the introduction, in the passive node model, the data associated with nodes is stored in a key-value store and clients operate on that data. Specifically, variables of node $k$ are stored as a pair $\br{k,v}$, where $v$ denotes  variables of node $k$. To execute an action of node $k$, a client (that is responsible for node $k$) reads the relevant values of variables required to perform the action and updates the relevant variables of node $k$.  

A program in the \pasnode model has a similar structure in terms of its nodes, variables, and actions, states, and transitions but differs the \actnode model in terms of the execution scheme. 
%
In \pasnode model, the system contains a set of clients. Each client is assigned (either statically or dynamically) a subset/partition of the whole set of nodes $V_p$. Each client is responsible for the execution of the actions of enabled nodes assigned to it. 
In an ideal environment, the execution of the program in \pasnode model is performed as follows: Let node $j$ be assigned to client $c1$. Then, $c1$ reads the values of the variables of $j$ and its neighbors. If it finds that some action of $j$ is enabled, it updates the key-value store with the new values for the variables of $j$. Similar to \actnode model, it is required that actions of multiple nodes can be serialized.

\textbf{Computation in the passive model. }
The notion of computation in passive-node model is identical to that of active-node model given above; the only difference is that we require clients to execute actions of each node assigned to them in a fair manner, where each node --that has an action enabled-- is executed infinitely often. 


\color{black}
\subsection{Stabilization}
\label{sec:stabilization}

In this section, we recall the definition of stabilization from \cite{EDW426}. Using the definition of computation from the previous section, stabilization is defined  as follows:


\textbf{Stabilization. }
Let $p$ be a program. Let $I$ be a subset of state space of $p$. We say that $p$ is stabilizing with state predicate $I$ iff 

\begin{description}
\item [Closure:] If program $p$ executes a transition in a state in $I$ then the resulting state is in $I$, i.e., for any transition $(s_0, s_1) \in \delta_p$, $s_0 \in I \Rightarrow s_1 \in I$, and 
\item [Convergence:] Any computation of $p$ eventually reaches a state in $I$, i.e., for any $\br{s_0, s_1, \cdots}$ that is a computation of $p$, there exists $l$ such that $s_l \in I$. 
\end{description}

A special case of stabilization is \textit{silent stabilization} where once the program reaches $I$ (denoted as the invariant of the program), it has no enabled actions, and, hence, the program will remain in that state forever (unless perturbed by faults). This paper focuses only on such \textit{silent stabilizing} programs. We refer the reader to \cite{NKD2019ICDCN} for discussion of non-silent stabilizing programs.  

%

\textbf{Silent Stabilization. }
Let $p$ be a program. Let $I$ be a subset of state space of $p$. We say that $p$ is silent stabilizing with state predicate $I$ iff 

\begin{description}
\item [Closure:] Program $p$ has no transitions that can execute in $I$, i.e., for any $s_0 \in I$, $(s_0, s_1) \not \in \delta_p$ for any state $s_1$, and \item [Convergence:] Any computation of $p$ eventually reaches a state in $I$, i.e., for any $\br{s_0, s_1, \cdots}$ that is a computation of $p$, there exists $l$ such that $s_l \in I$. 
\end{description}
\color{black}

\subsection{Voldemort Key Store}
\label{sec:voldemort-kv}

As discussed in Section \ref{sec:distprogram}, in the \pasnode model, the variables of all nodes are stored in a key-value store. In this paper, we use Voldemort --an open-source implementation of Amazon Dynamo \cite{DHJKLPSVV07SOSP}-- to implement the \pasnode model.

When a client wants to execute an action of the form $g \longrightarrow st$, it identifies all the variables required to execute this action. It issues a GET (read) request to all replicas (denoted by $N$, henceforth). It waits for receiving replies from at least $R$ --a configurable parameter in Voldemort-- replicas. If at least $R$ replicas reply before the timeout, the GET request is considered successful. If not, the client issues a second round of GET requests to the replicas. After the second, if replies are received from at least $R$ replicas in total (including the first round), the GET request is successful. Otherwise, it is not successful. If all reads are successful and the guard evaluates to true, the client identifies all variables that need to be changed. It then issues a PUT (write) request to all $N$ replicas. Similar to GET request, a PUT request is considered successful only if the client receives replies from at least $W$ --another configurable parameter in Voldemort-- replicas before timeout after at most two rounds. When the write is successful, action execution is complete. In the \pasnode model, the client does not have to retry an unsuccessful action. The clients learn the parameters $N$, $R$, and $W$ from the replicas at the time of startup. The above replication scheme employed by Voldemort is the active replication where the clients are in charge of data replication. The clients can also tune the values of $N$, $R$, $W$ if needed. By adjusting the value of $W$, $R$, and $N$, the consistency model of the key-value store is changed. For example, if $W + R > N$ and $W > \frac{N}{2}$ for every client, then then the consistency is sequential. If $W + R \le N$ then it is eventual consistency.
%
%

\subsection{Consistency Violating Faults (\cvf)}
\label{sec:cvf}
In the \pasnode model, the program state is stored at the replicas. In particular, for each variable $x$ of node $j$, each replica $i$ maintains a value of $x.j$ as a key-value pair. For the purpose of illustration, assume that there are three replicas and the values of $x.j$ at these replicas are $r_1, r_2$ and  $r_3$. Denote $f(r_1, r_2, r_3)$ as the abstract value of $x.j$ where $f$ is some resolution function that chooses a value among $r_1, r_2, r_3$ in a deterministic manner.
For example, function $f$ chooses the latest value of $x.j$ (assume that each value is also associated with a logical or physical timestamp).
%
In sequential consistency where the replicas work as if there is only a single replica, access (read/write) to variable $x.j$ by any client always returns the same abstract value of $x.j$. In eventual consistency, however, this property may be violated when different clients observe different values of $x.j$ (e.g. client $c1$ observes value $r_1$ while client $c2$ observes value $r_2$).

\textbf{Consistency Violating Faults (\cvf). } As the result of such violation in eventual consistency, the computation of the given program $p$ is a sequence $\br{s_0, s_1, \cdots}$ such that  most transitions $(s_l, s_{l+1}), l \geq 0$ in this sequence belong to $\delta_p$ (the set of transitions of $p$) and some transitions correspond to the scenario where some client working on node $j$ reads an inconsistent value for some variable and updates one or more variables of $j$. This incorrect transition is effectively the same as perturbing one or more variables of node $j$. We denote these incorrect transitions as concurrency violating faults ($\cvf_p$) and, by the above discussion, $\cvf_p$ is a \textit{subset} of 
%
$\{ (s_0, s_1) | s_0, s_1 \in S_p$ and $s_0$, $s_1$ differ only in the variables of some node $j$ of $p\}.$

\textbf{Remark. }
Whenever $p$ is clear from the context, we use $\cvf$ instead of $\cvf_p$. 

\textbf{Computation in the presence of \cvf. } With the introduction of \cvf, the computation of program $p$ in a eventually consistent replicated \pasnode model is of the form \br{s_0, s_1, \cdots} where

\begin{itemize}
\item  $\forall l: l \geq 0: $, $s_l$ is a state of $p$,
\item $\forall l: l \geq 0: (s_l, s_{l+1}) \in \delta_p \cup \cvf_p$ or\\ $(s_l = s_{l+1})$ and no action of $p$ is enabled in state $s_l$, and
\item If some action $ac$ of $p$ (of the form $g \longrightarrow st$) is continuously enabled (i.e., there exists $l$ such that $g$ is true in every state in the sequence after $s_l$) then $ac$ is eventually executed (i.e., for some $x \geq l$, $(s_x, s_{x+1})$ corresponds to execution of $st$.) 

\end{itemize}

\textbf{Stabilization of programs in the presence of \cvf. } \cvf occurs when data anomalies are introduced by eventual consistency, which is a rare scenario \cite{DHJKLPSVV07SOSP}. By design, \cvf is not deliberate and 

By design, \cvf is not deliberate and a single \cvf only perturbs the state of one node. \cvf also only occurs when data anomalies are introduced by eventual consistency, which is a rare scenario \cite{DHJKLPSVV07SOSP}. Thus, the program is likely to have the opportunity to execute several valid transitions between two \cvf transitions. Although some specific \cvf perturbations may significantly prolong the convergence of the program, the likelihood of such perturbations is small. Therefore, it is expected that a program in eventually consistent \pasnode model still stabilizes with an additional overhead for correcting \cvf in exchange for the higher performance of the weak consistency.

\subsection{Detect Rollback Approach}
\label{sec:detect-rollback}

In this section, we briefly recall the detect-rollback approach to handle \cvf. A fully detailed description of this approach is provided in \cite{NCKD2018LADC} (in the extended version), \cite{NCKD2018TR}, and \cite{NCKD2019JBCS}. In the detect-rollback approach, the user provides a correctness property $\Phi$ that the computation should always satisfy. We note that $\Phi$ can be the conjunction of smaller correctness properties, i.e. $\Phi = \bigcup_{i}\Phi_i$. The user runs the program on eventually consistent \pasnode model as well as the monitors. During the execution of program $p$, if property $\Phi$ is violated (any $\Phi_i = $ \si{false}) due to the occurrence of \cvf, the monitors will detect such violations and inform the computation to roll back to the most recent state where $\Phi$ is satisfied, and the computation is resumed from that state. In the detect-rollback approach, we assume each smaller predicate $\Phi_i$ is either a linear or semi-linear predicate as these predicates are common and can be detected efficiently \cite{CG98DC}. The monitors run predicate detection algorithms that are based on the algorithms by Chase and Garg \cite{CG98DC}. A more detailed description of the monitor algorithm is provided in \cite{NCKD2018LADC, NCKD2019JBCS}.
The execution of each action $ac = g \longrightarrow st$ at node $j$ is divided into two phases: (1) the read phase where the client read relevant program variables of $j$ and its neighbors (including securing exclusive access to these variables) to evaluate the guard $g$, and (2) the write phase where the client issues write requests to update one or more variables of $j$. We assume the read phase constitutes the majority of the time for action $ac$, and the detection latency of violations of $\Phi$ is significantly smaller than the read phase time (we observe that this assumption is valid from our experiments \cite{NCKD2018LADC}). With these assumptions, when a violation of $\Phi$ is detected, \textit{at least} one relevant client is in the read phase. Since the client in the read phase has not issued update requests, the action can be safely aborted without affecting the state of the computation. On the other hand, \textit{at most} one relevant client is in the write phase. Since the other competing client is in the read phase and will abort, the client in the write phase can safely finish the write phase without introducing conflicting data.
Based on this observation, the rollback algorithm works as follow: when a violation is detected by the monitors and reported to relevant clients, (i) if the client is in the read phase, it just aborts the current action and restart the execution of the action again, or (ii) if the client is in the write phase, it continues to finish the action normally.

\color{black}

\section{Experimental Evaluation and Analysis}
\label{sec:expriment}

In this section, we evaluate the rollback based approach and the stabilization-reliant approach to determine the benefits one can get if we use an eventual consistent key-value store instead of sequentially consistent key-value store. We proceed as follows: First, we identify the experimental setup to perform this comparison. Second, we identify the case studies that we use for this comparison. Next, we present and analyze the results of experiments performed in our local lab network. We also discuss some heuristics for improving the performance of the stabilization approach. Lastly, we present experiment results performed on Amazon AWS to confirm the local lab results.

\subsection{Experiment Setup}
\label{subsec:exp-setup}

\textbf{System configurations.}
We ran experiments on our local lab networks with 9 commodity PCs whose hardware configurations are specified in table \ref{table:machine-config}. 
Three of the PCs are dedicated to the Voldemort servers (replicas) and six other PCs are shared by the clients (each client machine hosts multiple Voldemort client programs). The number of servers is 3 and the number of clients is 30. With three servers, we choose N3R1W1 (the number of replicas N=3, the number of required reads R = 1 and the number of required writes W = 1) for eventual consistency, and N3R1W3 for sequential consistency since in our experiments N3R1W3 has better performance than N3R2W2. Hereafter, for brevity, we use R1W1 and R1W3 instead of N3R1W1 and N3R1W3 respectively.

We deployed the experiments on our local lab since we can control the environmental parameters such as network latency between the clients and servers. To adjust the network latency, we place a proxy process within each client machine that will relay all communication between the clients and the servers.
When a proxy is deployed for client $C$, it is co-located at the client machine. When $C$ wants to send a message to server $S$, it is immediately sent to the proxy. Proxy introduces the required delay before sending it to $S$. Communication from $S$ to $C$ is handled in a similar manner.  This allows us to evaluate the protocol in different network delay scenarios. 
A more detailed description of how the proxies work is provided in \cite{NCKD2018LADC, NCKD2019JBCS}.

Besides the experiments on the local lab network, we also ran experiments on Amazon Web Services (AWS) network to confirm the results in a more realistic environment. In AWS experiments, we used three EC2 M5.xlarge instances for the servers and six EC2 M5.large instances for the clients (cf. Table \ref{table:machine-config}). Delays involved on AWS are determined by the actual network conditions. 

\begin{table}[ht]
\caption{Configurations of machines used in the experiments}
\begin{center}
\begin{tabular}{|c|p{3cm}|l|l|l|}
    \hline
    Environment & Machine & CPU & RAM & Storage \\
    \hline
    \multirow{3}{*}{Local lab} & 3 server machines & 8 Intel Core i7-4770T 2.50 GHz  & 8 GB & SSD \\
    \cline{2-5}
    & 5 client machines & 4 Intel Core i5 660 3.33 GHz & 4 GB & HDD \\
    \cline{2-5}
    & 1 client machine  & 4 Intel Core i5-2500T 2.30GHz & 4 GB & HDD \\
    \hline
    \multirow{2}{*}{AWS} & 3 server machines (EC2 M5.xlarge) & 4 vCPUs  & 16 GB & SSD\\
    \cline{2-5}
    & 6 client machines (EC2 M5.large) & 2 vCPUs & 6 GB & SSD \\
    \hline
\end{tabular}
\label{table:machine-config}
\end{center}
\end{table}

\textbf{Workload partitioning schemes.} 
In the \pasnode model, each client is responsible for a (roughly equal) partition of the graph. We used three schemes to construct the clients' partitions. In the normal partitioning (or sequential partitioning), each client is responsible for trunk of a consecutive nodes (for example, client 0 is assigned the set of nodes from node 0 to node \SI{5,000}, client 1 is assigned from node \SI{5,001} to \SI{10,000}, and so on). In the Metis partitioning, we used graph partitioning tool Metis \cite{A-RK2006IPDPS} to partition the graphs. Metis partitioning algorithm aims to minimize the \textit{edge-cut} partitioning objective, i.e. the number of graph edges bridging different partitions, and thus increase the locality within the partitions. In other words, it helps reduce the amount of coordination between the clients. In the random partitioning, each client is assigned a distinct set with roughly the same number of nodes randomly selected from the graph.

\textbf{Termination detection algorithms.} Our termination detection algorithm to determine whether a program has reached a fixed point in the computation is based on the algorithm in \cite{DijkstraFG1983IPL}. For reason of space, we briefly describe the termination detection algorithm. Basically, the termination detector is also a Voldemort client program running a detection algorithm consisting of two rounds. In the first round, the algorithm reads the state of all nodes (including modification timestamps) and determines if every node has become disabled (i.e., all of its actions have the guards be evaluated to false). If that is true, it moves to the second round; otherwise, it restarts the first round.
In the second round, the algorithm checks if the state and modification timestamp of every node is unchanged since the most recent first-round check. If there is any change, the algorithm restarts from the first round; otherwise, it reports the termination of computation. 
The termination detector runs in the consistency mode where $R = N$ (the number of required reads equals the number of replicas) to ensure the reliability.
Since the termination detector only reads from the key-value store, it minimally affects the stabilization time of the computation.

\textbf{Client execution modes.} The clients were configured to run in four different modes corresponding to four different ways of executing the computation. 
In \textit{sequential} mode (SEQ), the clients run on sequentially consistent key-value store and use mechanisms (e.g. locks) to guarantee exclusive access to the data. No \cvf should occur in sequential mode. This is the standard approach for executing the computation and is used as the baseline for comparison.
In \textit{eventual with stabilization} mode (EVE-S), the clients also employ mutual exclusion mechanisms but run on eventually consistent data store. This mode allows \cvf to occur due to eventual consistency. However, \cvf is expected to be infrequent so that between two instances of \cvf, the clients can execute several transitions to stabilize the computation. This mode is the typical way to implement self-stabilization approach.
\textit{Eventual with aggressive stabilization} mode (EVE-AS) is similar to eventual with stabilization except that the clients do not use mutual exclusion mechanisms for exclusive access of the data. Consequently, more \cvf are expected in this mode. This mode is a more aggressive way of self-stabilization approach.
Lastly, in \textit{rollback} mode, the clients run on eventually consistent data store and also use mutual exclusion mechanisms. Hence, \cvf occurs in rollback mode. However, instead of relying on the stabilizing transitions of the program to correct \cvf, the monitors are deployed to detect violations and the computation is then rolled back to undo the effect of \cvf. This mode represents the detect-rollback approach.

\subsection{Case Study Problems}
\label{subsec:exp-case-studies}
We used three self-stabilization problems as our case studies: planar graph coloring, arbitrary/general graph coloring, and maximal matching.
The problem of planar graph coloring is motivated by applications on planar graph such as weather monitoring \cite{FBMA2013PCS}, radio-coloring in wireless and sensor network \cite{PSNS1995PODC}, computing Voronoi diagram \cite{NXIA2008CCCG}, etc.. To color a planar graph, we implemented the self-stabilizing planar coloring algorithm by Ghosh and Karaata \cite{GK93DC} that guarantees to use at most 6 colors. 
For arbitrary/general graph coloring, we used the self-stabilizing algorithm by Gradinariu and Tixeuil \cite{GT2000OPODIS} (the first of three variations). The problem of coloring a general graph has many uses in classical applications such as scheduling, resource allocation, pre-processing the graph, to more recent applications in banking and financial services \cite{GLG2017RecSys}, social network analysis \cite{RA2014SNAM}. 
The problem of matching also has many resource-allocation-based applications such as telephone line switching \cite{HM1977SECCGC}, college student placement \cite{BS1999JET}, stable marriage \cite{knuth1997stablemarriage}, and matrix computation \cite{PF1990TMS}. We used the self-stabilizing algorithm by Manne et al. \cite{MMPT2009TCS} to find the maximal matching of a graph.

We used three types of input graphs in the experiments: planar graphs, social network graphs, and random regular graphs. A planar graph is a graph that can be drawn on a plane such its edges do not cross with each other. To generate planar graphs, we used the uniform random sampling algorithm and program by Eric Fusy \cite{Fusy2009RSA}. The program in \cite{Fusy2009RSA} generates an arbitrary planar graph. We chose to use it to generate a graph with approximately 10,000 nodes. However, since this program cannot be tuned to get a graph with exactly 10,000 nodes, the graph we use is one with 11,033 nodes and 24,333 edges.
A social network graph has the degrees of its nodes follow the power-law distribution and the nodes form clusters within the graph. A random regular graph is a graph whose nodes have the same degree (in our experiment, each node has 6 neighbors) and the edges are randomly distributed among the nodes. Regular graphs have the advantage that the workload is evenly distributed between the clients and the interaction between clients is random.
We used the tool \textit{networkx} \cite{networkx} to generate social network graphs and regular graphs. These graphs have 10,000 to 50,000 nodes.

\subsection{Experiment Results}
\label{subsec:exp-result}

In section \ref{subsec:result-basic:ss-vs-rollback}, we describe experimental results of running the two approaches (self-stabilization and detect-rollback) on the test cases in our local lab network. We also investigate factors affecting the performance. Sections \ref{subsec:result-extra-var} describes some heuristics to improve the stabilization time of the computation.
Finally, Section \ref{subsec:result-aws} presents experimental results on Amazon AWS network.

\subsubsection{Benefit of Self-stabilization vs. Detect-rollback: Comparison and Analysis}
\label{subsec:result-basic:ss-vs-rollback}

Table \ref{tab:self-stabilization-vs-rollback} shows the experiment results of running four execution modes (cf. Section \ref{subsec:exp-setup}, client execution modes) on different types of problems and input data. The sequential mode (SEQ) is used as the baseline of comparison. The stabilization approach is represented by two modes: eventual with stabilization (EVE-S) and eventual with aggressive stabilization (EVE-AS). The detect-rollback approach is represented by rollback mode. 

\begin{table}[htbp]
\caption{The convergence benefit of stabilization vs. detect-rollback in different types of problems and input data. Input graph is partitioned in the normal/sequential scheme. Sequential mode (SEQ) is the baseline for comparison. Stabilization approach includes two execution modes: eventual with stabilization (EVE-S) and eventual with aggressive stabilization (EVE-AS). Detect-rollback approach is represented by rollback mode (cf. Section \ref{subsec:exp-setup}).
Rows 7-10 are convergence time benefits, shown in percentage increase or in speedup (e.g. $\times 5.2$ means 5.2 times faster).}
\begin{tabular}{|p{1.5cm}|p{2.8cm}|p{1.3cm}|p{1.2cm}|p{1.2cm}|p{1.1cm}|p{1.1cm}|p{1.1cm}|} 
\hline
& {\parbox{1.5cm}{Problem}} & \multicolumn{1}{c|}{\parbox{1.3cm}{Planar Graph Coloring}} & \multicolumn{2}{c|}{\parbox{2.4cm}{Arbitrary Graph \\Coloring}} & \multicolumn{3}{c|}{\parbox{3.6cm}{Maximal Matching}} \\ \hline
& Input graph & Planar 10K & Social 50K & Regular 50K & Social 10K & Regular 10K & Planar 10K\\ 
\hline
\multirow{4}{*}{\parbox{1.5cm}{\centering{Conver-gence time (seconds)}}} & SEQ                   & 3,887 & 27,995 & 6,518 & 31,581 & 14,859 & 8,545 \\ 
& EVE-S                   & 2,658 & 18,229 & 4,270 & 23,246 & 11,028 & 6,173\\ 
& EVE-AS                 & 754  &  1,885 & 3,547 & 2,892 & 1,866 & 2,590 \\ 
& Rollback          & 3,860 & 32,165 & 4,624 & 32,238 & 12,496 & 8,660 \\  
\hline
\multirow{4}{*}{\parbox{1.5cm}{\centering{Benefit}}}& EVE-S vs. SEQ       & 31.6\% & 34.9\% & 34.5\% & 26.4\% & 25.8\% & 27.8\%  \\ 
& EVE-AS vs. EVE-S     & $\times 3.5$ & $\times 9.7$ & $\times 1.2$ & $\times 8$ & $\times 5.9$ & $\times 2.4$\\ 
& EVE-AS vs. SEQ     & $\times 5.2$ & $\times 14.9$ & $\times 1.8$ & $\times 10.9$ & $\times 8$ & $\times 3.3$ \\ 
& Rollback vs. SEQ & 0.7\% & -14.9\% & 29\% & -2.1\% & 15.9\% & -1.4\% \\ 
\hline
\end{tabular}
\label{tab:self-stabilization-vs-rollback}
\end{table}

We observe that the benefit of an approach in the same problem depends on the type of input graph. Consider using the detect-rollback approach for the maximal matching problem. This approach is worse than the baseline in social graphs, comparable in planar graphs, and better in regular graphs. We anticipate that one of the reason is related to the structure of input graphs. 

\begin{figure}
    \begin{subfigure}[t]{0.45\textwidth}
        \centering
        \includegraphics[width=1\textwidth]{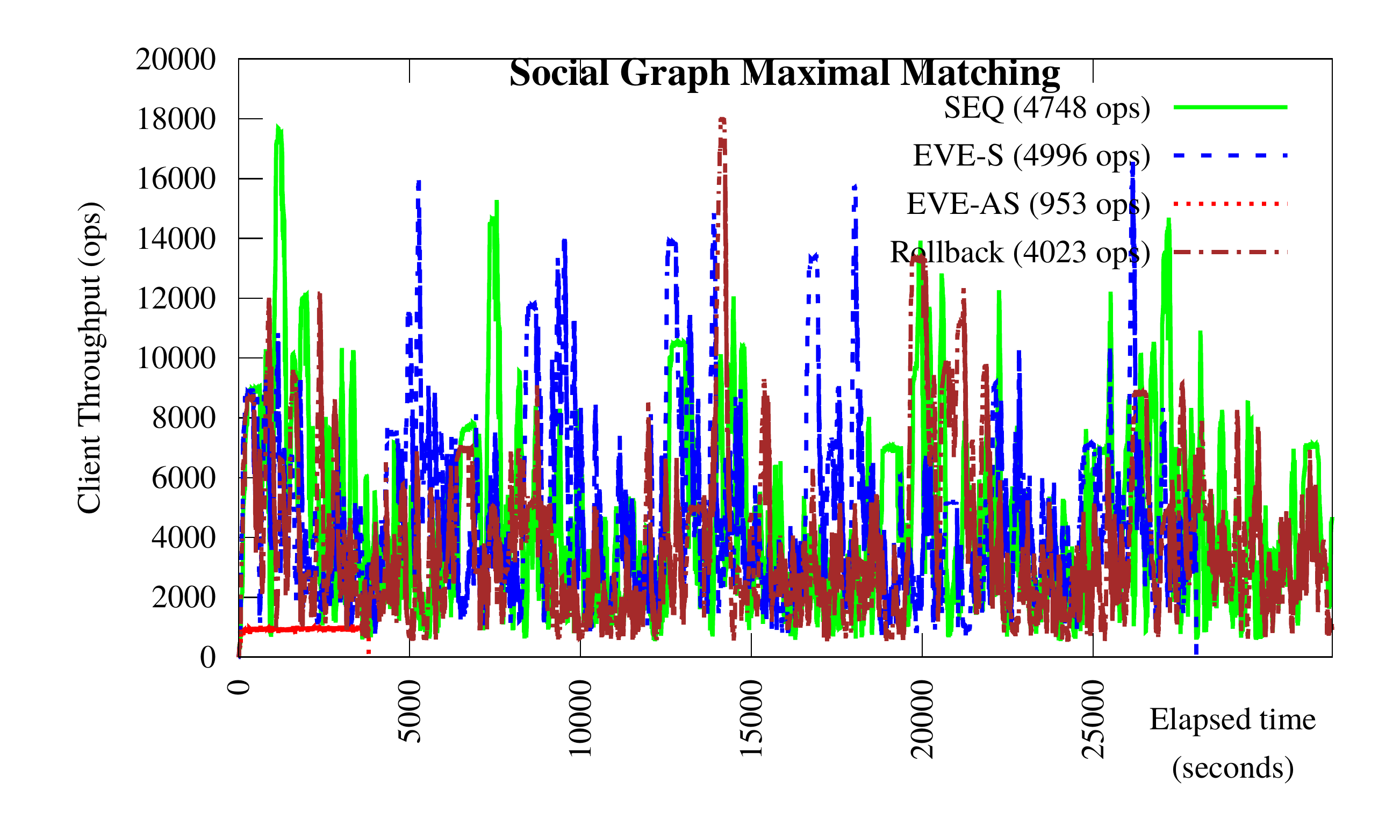}
        \caption{Social graph}
        \label{fig:max-matching-throughput-powerlaw}
    \end{subfigure}\hfill
    \begin{subfigure}[t]{0.45\textwidth}
        \centering
        \includegraphics[width=1\textwidth]{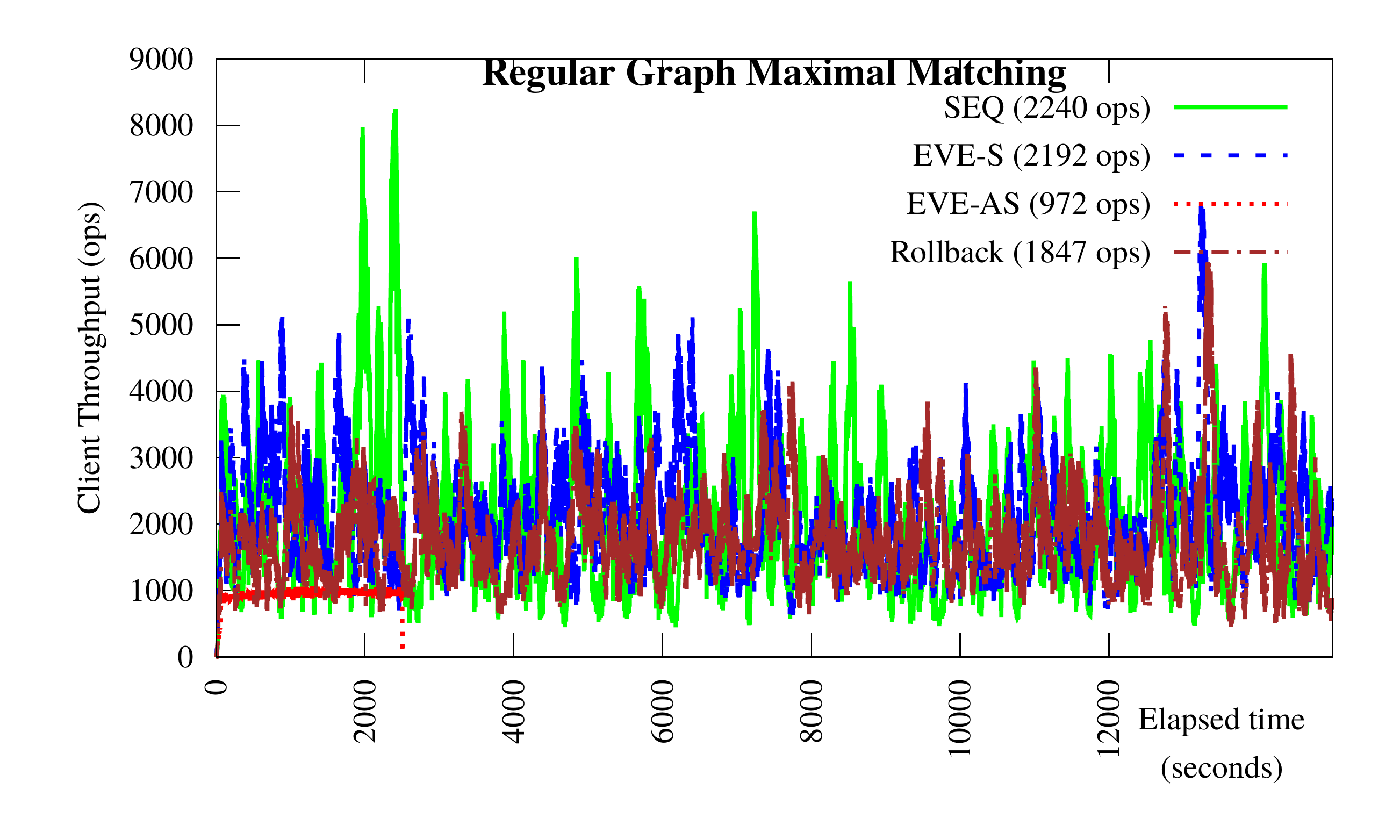}
        \caption{Regular graph}
        \label{fig:max-matching-throughput-regular}
    \end{subfigure}\\
    \begin{subfigure}[t]{0.45\textwidth}
        \centering
        \includegraphics[width=1\textwidth]{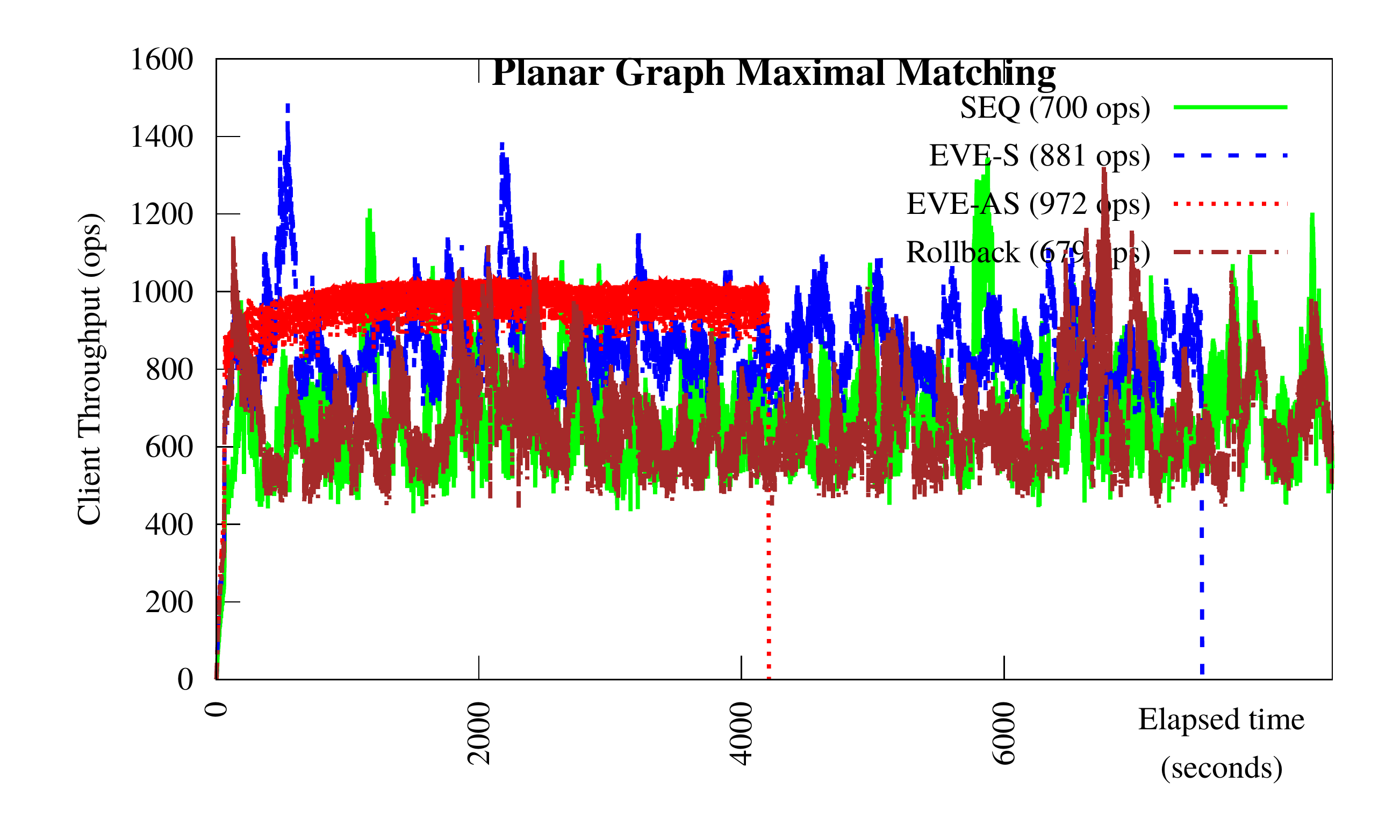}
        \caption{Planar Graph}
        \label{fig:max-matching-throughput-planar}
    \end{subfigure}\hfill
    \begin{subfigure}[t]{0.4\textwidth}
        \centering
        \includegraphics[width=0.9\textwidth]{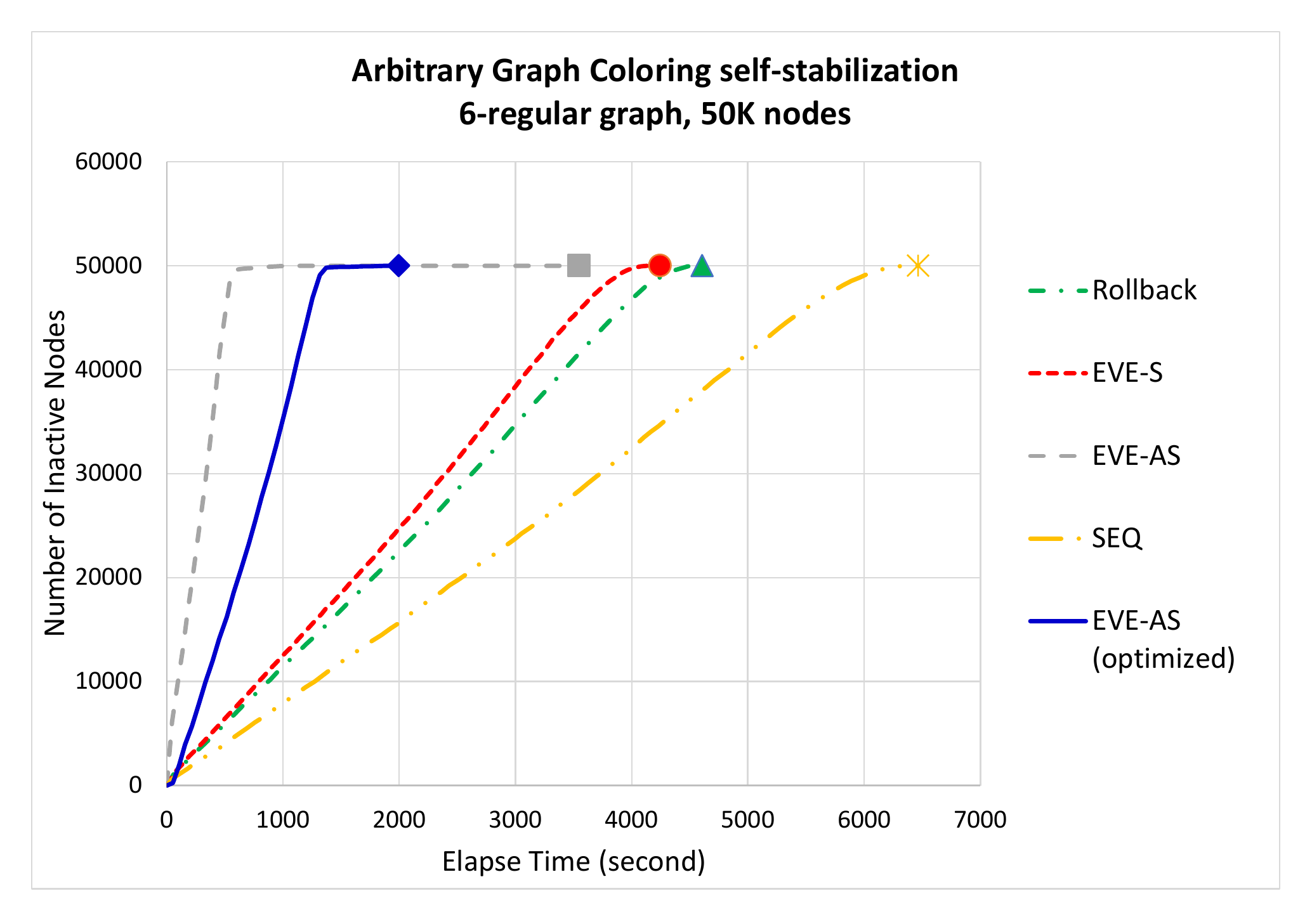}
        \caption{Convergence pattern}
        \label{fig:arbitrary-coloring-convergence-graph}
    \end{subfigure}
\caption{Sub-figures (a), (b), (c): Measurement of client throughput (in ops -- operations per seconds) of maximal matching program with different input graphs. Sub-figure (d): The convergence pattern of different execution modes in the arbitrary graph coloring problem.}
\label{fig:max-matching-throughput}
\end{figure}

\textbf{Impact of input graph structure.} In social graphs, there are a few nodes with a very high number of neighbors. When a client working on such a node, in order to make sure no other client is working on one of this node's neighbors, the client is likely to have to wait for a significant amount of time to obtain the exclusive access. As shown in Figure \ref{fig:max-matching-throughput-powerlaw}, the throughput of all execution mode where mutual exclusion is employed (SEQ, EVE-S, and rollback) is much higher than EVE-AS mode where mutual exclusion is eliminated. We note that the throughput is not related to the convergence time but useful to understand the behavior of the program. Specifically, when the difference in throughput of execution modes with mutual exclusion and EVE-AS is very \textit{high} and \textit{consistent}, we anticipate that the overhead of mutual exclusion is a major contributor to that difference.
For example, by comparing the average throughput of EVE-S and EVE-AS, we estimate that about 80\% of execution time in EVE-S mode is spent for securing exclusive access to data items.
By contrast, in regular graph, only about 50\% of the time is spent for such job (cf. Figure \ref{fig:max-matching-throughput-regular}). The reason for this reduction is that in regular graphs, each node has roughly the same degree, the edges are randomly distributed among the nodes. Therefore, comparing to social graphs, the chance of two clients working on two neighboring nodes and the amount of time a client spend to obtain exclusive access for a node is smaller in regular graphs. A smaller mutual exclusion overhead also implies a smaller chance of conflicts between clients, a fewer number of rollbacks in the rollback mode, and, in case the clients roll back, a smaller the amount of work wasted.
Another factor of consideration is that the client workload is not even in social graphs: clients assigned with high degree nodes have more work to do, thus the convergence time of the whole computation is determined by the convergence time of (the partitions associated with) those clients. As previously explained, these clients suffer from a higher chance of conflicts and rollbacks. Consequently, for social graphs, the convergence time of rollback mode is extended longer.
It is possible to improve the performance of the rollback mode by finding and processing high degree nodes in advance. These nodes are distinct and there is only a small number of them in the graph. However, we omitted this option in this paper since this structural information is not always available.

In planar graphs, as suggested by Figure \ref{fig:max-matching-throughput-planar}, the overhead for mutual exclusion is small. In contrast to Figures \ref{fig:max-matching-throughput-powerlaw} and \ref{fig:max-matching-throughput-regular}, we observe the throughput of EVE-AS is higher than execution modes with mutual exclusion (SEQ, EVE-S, and rollback). In this case, the throughput difference does not indicate the overhead of mutual exclusion. However, by comparing with Figures \ref{fig:max-matching-throughput-powerlaw} and \ref{fig:max-matching-throughput-regular}, we observe the throughput for planar graphs is much smaller. Thus, the overhead of mutual exclusion is small. Consequently, the chance of client conflicts and rollbacks is small. This is because we can partition a planar graph into almost non-overlapping partitions. When working on internal nodes, clients already have exclusive access. Mutual exclusion mechanism is used only for nodes at the border. Nevertheless, the rollback mode is still not better than sequential mode because of the skewed client workload in the normal partition. If we partition planar graphs with random partitions, the workload is more even and the convergence time of rollback mode is $22.3\%$ faster than convergence time of sequential mode. Because of space reason, the results of random partitioning is presented in Appendix \ref{sec:appendix-random-partition-planar}.

With respect to the benefit of stabilization approach (EVE-S and EVE-AS modes), we observe that the benefit of EVE-S mode is relatively stable within the range 25\%--35\%, when the problem and input graph is changed. This is because the benefit of EVE-S stems from the difference between eventual consistency and sequential consistency of the data store. On the other hand, the benefit of EVE-AS mode depends on the input graphs. If the overhead of mutual exclusion is high, the benefit of EVE-AS is high, and vice versa. For example, the convergence time of EVE-AS is 11 times faster than SEQ on social graphs (where there is the most mutual exclusion overhead), and is only 3 times faster than SEQ in planar graphs (where there is the least mutual exclusion overhead). We anticipate the reason for this observation is that in EVE-AS mode, the mutual exclusion overhead is completed removed. Although some additional \cvf are introduced, the benefit of removing mutual exclusion overhead outweighs the cost of correcting additional \cvf.

\textbf{Impact of case study problems.} 
For the same type of input graph, for example, planar graph, the benefit of the eventual with aggressive stabilization (EVE-AS) is higher in the planar graph coloring problem than in maximal matching problem. If the input graph is a social graph, the benefit in the graph coloring problem is higher than in maximal matching. We anticipate the reason is related to the effect of \cvf{s} and the cost of correct them. Specifically, in coloring related problems, \cvf{s} can cause a client to assign a color that is the same as one of its neighbors' colors. However, this violation can be fixed by one program action at the conflicting neighbor. Nodes at a distance of more than 1 hop are not affected by the error. By contrast, in maximal matching problem, an inconsistent matching can have a cascading effect that requires updates at distant nodes. As an illustration, suppose we have four nodes $v_1$, $v_2$, $v_3$, and $v_4$ on a straight line in that order. Suppose both $v_1$ and $v_3$ are matched with $v_2$ (due to \cvf). To correct this error, we can unmatch $v_2$ from $v_3$. Since $v_3$ is now free, it can be matched with the free node $v_4$, thus updating the states of both $v_3$ and $v_4$. Therefore, the cost for correcting \cvf{s} in maximal matching problem is usually higher and the benefit of EVE-AS is smaller. We also note an exception in the experimental results. The benefit of graph coloring is smaller than maximal matching on regular graphs. Further investigation shows that eliminating mutual exclusion mechanisms in graph coloring can cause some \cvf that is difficult to recover. For example, suppose two clients $C_1$ and $C_2$ are working on two nodes $v_1$ and $v_2$ at the same time. Suppose the original color of both nodes is 0. Because no mutual exclusion is used, both clients may assign the same new colors 1 for both nodes, causing another violation. This error is usually resolved when one of the clients visits its node in the next round and change its node to a different color. However, if both $C_1$ and $C_2$ re-visit $v_1$ and $v_2$ at the same time, the problem may continue to persist. This scenario is unlikely to happen in a social graph because the client workload is not even. However, in a regular graph, because the client workload is fairly evenly distributed, the conflicting color may take a long time to be resolved. In other words, running graph coloring in EVE-AS mode does not guarantee convergence. One possibility to address this problem is to modify the coloring algorithm so that the client would choose a random value among available colors for its nodes. With this modification, EVE-AS is probabilistic self-stabilizing. In our experiments with the random coloring scheme, the convergence time of coloring the same regular graph in EVE-AS mode improves from \SI{3,547}{seconds} to \SI{1,431}{seconds}. On the other hand, the convergence time for social graph stays almost the same (cf. Table \ref{tab:stabilization-optimization}).

\subsubsection{Improving the convergence time in stabilization approach}
\label{subsec:result-extra-var}

In the passive node model, clients are responsible for checking which nodes have enabled actions and execute those actions. In Table \ref{tab:self-stabilization-vs-rollback}, we considered the case where clients evaluated the guards of nodes assigned to them in a round-robin manner. One of the issues with this is that some nodes whose actions are enabled may not be considered while the client is evaluating other nodes assigned to it but have no enabled actions. Note that this issue is ignored in the active node model, as, generally, it is assumed that the scheduler will choose some active node for execution. The time required to determine this node is ignored. 

In the programs under consideration, if no action of $j$ is enabled in the current state then this information is stable until a neighbor of $j$ executes. Thus, if a node could tell the client that its actions are unlikely to be enabled then the client can save on reading the states of its neighbors. For such an approach to work, for node $j$, we need to know (1) $nd\_change.j$  the last time the client checked that no actions are enabled at $j$, and (2) $nbr\_change.j$ the last time one of its neighbors was updated. 

Thus, when client reads the state of $j$ and finds that $nd\_change.j > nbr\_change.j$, it does not need to read the state of its neighbors to determine if some action of $j$ is enabled. Since clocks of all computers involved may not be identical, we change this to $nd\_change.j > nbr\_change.j + \Delta.j + \epsilon$ where $\Delta.j$ is the length of the last execution of $j$ and $\epsilon$ is the upper bound for clock synchronization error. In other words, if $nd\_change.j > nbr\_change.j + \Delta + \epsilon$  true then the client can save time by not issuing GET requests to neighbors of $j$. 

Table \ref{tab:stabilization-optimization} considers execution with this optimization. We find that this optimization is useful only when the convergence pattern exhibits a long tail at the end (cf. EVE-AS mode in Figure \ref{fig:arbitrary-coloring-convergence-graph}). 
The overhead of the optimization (for reading and writing additional variables) causes EVE-AS (optimized) converge slower than EVE-AS at first. However, the optimization significantly reduces the tail of convergence graph and thus improves the overall convergence time by 44\%. If the convergence pattern does not have the long tail characteristic (such as EVE-S mode in Figure \ref{fig:arbitrary-coloring-convergence-graph}, or EVE-AS mode with random coloring), this optimization increases the convergence time because of the extra overhead.

\begin{table}[htbp]
\centering
\caption{Effectiveness of the random coloring and the optimization for stabilization approach in the arbitrary graph coloring problem. Convergence time is measured in seconds.}
\begin{tabular}{|p{2cm}|p{2cm}|p{2cm}|p{2cm}|p{2cm}|} 
\hline
Execution mode & Optimization & New color & Regular graph 50K & Social graph 50K \\
\hline
EVE-AS & Yes   & Deterministic & 1,972 & 4,805 \\
EVE-AS & Yes   & Random        & 1,941 & 4,807 \\
EVE-AS & No  & Deterministic & 3,547 & 1,885 \\
EVE-AS & No  & Random        & 1,431 & 1,883 \\
\hline
EVE-S  & No   & Deterministic & 4,270 & 18,229 \\ 
EVE-S  & Yes   & Deterministic & 5,136 & $> 20,000$ \\
\hline
\end{tabular}
\label{tab:stabilization-optimization}
\end{table}

\subsubsection{Experiments on Amazon AWS}
\label{subsec:result-aws}
To confirm the results in a more realistic deployment, we also run experiments on Amazon Web Services (AWS) network. As shown in Table \ref{tab:aws-self-stabilization-vs-rollback}, the AWS results generally agree with the experimental results on our local lab.

\begin{table}[htbp]
\caption{Experiment results on Amazon AWS network.}
\begin{tabular}{|p{1.5cm}|p{2.8cm}|p{2cm}|p{2cm}|p{2cm}|} 
\hline
& {\parbox{1.5cm}{Problem}} & \multicolumn{1}{c|}{\parbox{2cm}{Planar Graph Coloring}} & \multicolumn{1}{c|}{\parbox{2cm}{Arbitrary Graph \\Coloring}} & \multicolumn{1}{c|}{\parbox{2cm}{Maximal Matching}} \\ \hline
& Input graph & Planar 10K & Social 10K & Regular 10K \\ \hline
& Partition scheme & Random & Sequential & Sequential \\ 
\hline
\multirow{4}{*}{\parbox{1.5cm}{\centering{Conver-gence time (seconds)}}} & SEQ                   & 10,211 & 21,265 & 6,816 \\ 
& EVE-S                  & 6,586 & 13,630 & 4,038 \\ 
& EVE-AS                 & 797 & 2,430 & 413 \\ 
& Rollback          & 9,575 & 21,718 & 7,625 \\  
\hline
\multirow{4}{*}{\parbox{1.5cm}{\centering{Benefit}}}& EVE-S vs. SEQ       & 35.5\% & 35.9\% & 41.7\% \\ 
& EVE-AS vs. EVE-S     & $\times 8.3$ & $\times 5.6$ & $\times 9.8$ \\ 
& EVE-AS vs. SEQ     & $\times 12.8$ & $\times 8.8$ & $\times 16.5$ \\ 
& Rollback vs. SEQ   & 6.2\% & -2.1\% & -11.9\% \\ 
\hline
\end{tabular}
\label{tab:aws-self-stabilization-vs-rollback}
\end{table}

\section{Conclusion}
\label{sec:concl}

In this paper, we considered the passive node model introduced in \cite{NKD2019ICDCN} and two approaches to reduce the time for convergence in it. Specifically, in the passive node model, the data associated with nodes is stored in a key-value store. If we use sequential consistency (with mutual exclusion) then execution of the program is consistent. However, sequential consistency can reduce performance when compared with weaker eventual consistency. We considered the effect of dealing with such inconsistency -- denoted consistency violation faults (\cvf{s}). The first relied on detecting them and rolling back. The second, applicable only to stabilizing programs, was to observe that \cvf{s} are a subset of transient faults and, hence, are already tolerated. 
Our analysis shows that for stabilizing programs, the second approach provides substantial benefits compared with the first one. Specifically, the second approach provides a 25\%--35\% improvement for different applications. Especially, in the aggressive stabilization mode that removes mutual exclusion and treats mutual exclusion violations as additional \cvf{s}, the convergence time is improved by 2--15 times. 
By contrast, the rollback based approach provides limited benefit and potentially causes performance to suffer when compared with sequential consistency. 

We also considered another approach to reduce the time for convergence. It relied on a heuristics to allow clients to keep track of nodes which may have enabled actions. Experimental results show that the heuristics can improve convergence time by 44\%. However, this approach potentially loses stabilization property if heuristics cause a client to incorrectly think that one of the nodes assigned to it does not have an enabled action. Furthermore, it is only suitable for convergence patterns with a long tail of slow progress at the end.

We also find that the stabilization based approach can benefit even more if the application can use other techniques to reduce overall time. Specifically, we considered the use of graph partitioning to reduce \cvf{s}. In this case, both approaches showed benefits. But the benefit of stabilization-based approach was higher. 

From the analysis of this work, we find that stabilization based approach provides a substantial benefit compared with rollback based approach. However, in both cases, the time required for convergence of the last few nodes is still quite high. One of the future work in this area is to reduce this overhead.


\section*{Acknowledgments}
This work is supported by NSF XPS 1533802.

\bibliography{DuongNguyen}

\newpage
\appendix
\section{Appendix}
\label{sec:appendix}

\subsection{Improving the Benefit of Detect-Rollback in Planar Graph with Random Partitioning}
\label{sec:appendix-random-partition-planar}

We observe that for both planar coloring and maximal matching problems, the convergence time of rollback mode is roughly the same as SEQ mode.
The reason is that with the normal/sequential partitioning scheme, each client is assigned a consecutive chunk of graph nodes. However, the graph edges are not evenly distributed in the graph, with nodes of lower identifiers have more edges (cf. Table \ref{tab:planar-normal-partition}). This will cause some imbalance in workload between clients as well as clients assigned with lower ID nodes are more likely to conflict with each other, especially when the clients are working on the first nodes of their assigned partitions. When conflict occurs, these clients have to roll back and redo conflicting tasks. Furthermore, frequent conflicts at the early time of the execution cause the clients to switch to sequential consistency early without taking advantage of eventual consistency. This issue can be improved if we partition the graph so that the workload and the edges are more evenly distributed between clients (cf. Table \ref{tab:planar-random-partition}). Table \ref{tab:self-stabilization-vs-rollback-planar-random-partition} compare experimental results when the planar graph is partitioned in sequentially and randomly. We observe that the benefit of detect-rollback is significantly improved. Moreover, the speedup benefit of stabilization is also improved.


\begin{table}[htbp]
\centering
\caption{Benefit of Self-stabilization vs. Detect-rollback in maximal matching and planar coloring problems. Input graph is a planar graph with normal partitions and random partitions. Rows 2-5 are stabilization time (in seconds). Rows 6-8 are stabilization time benefits, shown in percentage increase or in speedup (e.g. $\times 3$ means 3 times faster).}
\begin{tabular}{|p{3cm}|p{2cm}|p{1.6cm}|p{1.6cm}|p{1.6cm}|p{1.6cm}|} 
\hline
& \multirow{2}{*}{\parbox{2cm}{Execution mode}} & \multicolumn{2}{c|}{Maximal Matching} & \multicolumn{2}{c|}{Maximal Matching} \\
\cline{3-6}
&  & Normal partition & Random partition & Normal partition & Random partition \\
\hline
\multirow{4}{*}{\parbox{3cm}{\centering{Convergence time (seconds)}}} & SEQ & 8,545 & 10,736 & 3,887 & 8,686\\ 
& EVE-S                   & 6,173 & 7,026 & 2,658 & 5,315 \\ 
& EVE-AS                  & 2,590 & 1,448 & 754 & 655 \\ 
& Rollback                & 8,660 & 8,341 & 3,860 & 7,242 \\  
\hline
\multirow{4}{*}{\parbox{3.5cm}{\centering{Benefit}}} & EVE-S vs. SEQ    & 27.8\% & 34.6\% & 31.6\% & 38.8\% \\ 
& EVE-AS vs. SEQ       & $\times 3.3$ & $\times 9.4$ & $\times 5.2$ & $\times 13.3$ \\ 
& Rollback vs. SEQ     & -1.4\% & 22.3\% & 0.7\% & 16.6\% \\ 
\hline
\end{tabular}
\label{tab:self-stabilization-vs-rollback-planar-random-partition}
\end{table}

\begin{table}[htbp]
\caption{Some properties of planar graph with normal partitioning scheme.}
\begin{tabular}{|p{1cm}|p{1cm}|p{1cm}|p{1cm}|p{1cm}|p{1cm}|p{1cm}|p{1cm}|p{1cm}|}
\hline 
{\parbox{1cm}{Partition scheme}} & {\parbox{1cm}{Partition Id}} & {\parbox{1cm}{max Degree}} & {\parbox{1cm}{Min Degree}} & {\parbox{1cm}{Total Degree}}& {\parbox{1cm}{node Count}}& {\parbox{1cm}{Avg Degree}} & {\parbox{1cm}{external Edges}} & {\parbox{1cm}{internal Edges}} \\
\hline
\multirow{30}{*}{\parbox{1cm}{\centering{Normal}}} & 0 & 22 & 3 & 2058 & 370 & 5.56 & 476 & 1582 \\
 & 1 & 16 & 3 & 1972 & 370 & 5.33 & 628 & 1344 \\
 & 2 & 22 & 3 & 1886 & 370 & 5.1 & 486 & 1400 \\
 & 3 & 22 & 3 & 2029 & 370 & 5.48 & 679 & 1350 \\
 & 4 & 18 & 3 & 2041 & 370 & 5.52 & 789 & 1252 \\
 & 5 & 21 & 3 & 2031 & 370 & 5.49 & 775 & 1256 \\
 & 6 & 20 & 3 & 2090 & 370 & 5.65 & 946 & 1144 \\
 & 7 & 14 & 3 & 1891 & 370 & 5.11 & 317 & 1574 \\
 & 8 & 17 & 3 & 1849 & 370 & 5 & 685 & 1164 \\
 & 9 & 17 & 3 & 2038 & 370 & 5.51 & 582 & 1456 \\
 & 10 & 16 & 3 & 1887 & 370 & 5.1 & 473 & 1414 \\
 & 11 & 14 & 3 & 1784 & 370 & 4.82 & 530 & 1254 \\
 & 12 & 15 & 3 & 1867 & 370 & 5.05 & 343 & 1524 \\
 & 13 & 22 & 3 & 1944 & 370 & 5.25 & 900 & 1044 \\
 & 14 & 16 & 3 & 1958 & 370 & 5.29 & 652 & 1306 \\
 & 15 & 19 & 3 & 1968 & 370 & 5.32 & 560 & 1408 \\
 & 16 & 17 & 3 & 1904 & 370 & 5.15 & 436 & 1468 \\
 & 17 & 18 & 3 & 1761 & 370 & 4.76 & 577 & 1184 \\
 & 18 & 18 & 3 & 1922 & 370 & 5.19 & 472 & 1450 \\
 & 19 & 16 & 3 & 1791 & 370 & 4.84 & 507 & 1284 \\
 & 20 & 14 & 3 & 1674 & 370 & 4.52 & 596 & 1078 \\
 & 21 & 16 & 3 & 1748 & 370 & 4.72 & 492 & 1256 \\
 & 22 & 14 & 2 & 1059 & 370 & 2.86 & 637 & 422 \\
 & 23 & 10 & 2 & 927 & 370 & 2.51 & 617 & 310 \\
 & 24 & 6 & 2 & 844 & 360 & 2.34 & 638 & 206 \\
 & 25 & 6 & 2 & 840 & 360 & 2.33 & 652 & 188 \\
 & 26 & 8 & 2 & 850 & 360 & 2.36 & 628 & 222 \\
 & 27 & 10 & 2 & 855 & 360 & 2.38 & 651 & 204 \\
 & 28 & 10 & 1 & 785 & 360 & 2.18 & 529 & 256 \\
 & 29 & 4 & 1 & 413 & 353 & 1.17 & 317 & 96 \\
\hline
\end{tabular}
\label{tab:planar-normal-partition}
\end{table}

\begin{table}[htbp]
\caption{Some properties of planar graph with random partitioning scheme.}
\begin{tabular}{|p{1cm}|p{1cm}|p{1cm}|p{1cm}|p{1cm}|p{1cm}|p{1cm}|p{1cm}|p{1cm}|}
\hline 
{\parbox{1cm}{Partition scheme}} & {\parbox{1cm}{Partition Id}} & {\parbox{1cm}{max Degree}} & {\parbox{1cm}{Min Degree}} & {\parbox{1cm}{Total Degree}}& {\parbox{1cm}{node Count}}& {\parbox{1cm}{Avg Degree}} & {\parbox{1cm}{external Edges}} & {\parbox{1cm}{internal Edges}} \\
\hline
\multirow{30}{*}{\parbox{1cm}{\centering{Random}}} &  0 & 19 & 1 & 1606 & 370 & 4.34 & 1540 & 66 \\
 & 1 & 17 & 1 & 1659 & 370 & 4.48 & 1609 & 50 \\
 & 2 & 15 & 1 & 1620 & 370 & 4.38 & 1570 & 50 \\
 & 3 & 18 & 1 & 1721 & 370 & 4.65 & 1659 & 62 \\
 & 4 & 14 & 1 & 1559 & 370 & 4.21 & 1513 & 46 \\
 & 5 & 18 & 1 & 1595 & 370 & 4.31 & 1543 & 52 \\
 & 6 & 16 & 1 & 1754 & 370 & 4.74 & 1668 & 86 \\
 & 7 & 13 & 1 & 1635 & 370 & 4.42 & 1569 & 66 \\
 & 8 & 18 & 1 & 1726 & 370 & 4.66 & 1682 & 44 \\
 & 9 & 21 & 1 & 1681 & 370 & 4.54 & 1625 & 56 \\
 & 10 & 14 & 1 & 1618 & 370 & 4.37 & 1576 & 42 \\
 & 11 & 22 & 1 & 1613 & 370 & 4.36 & 1561 & 52 \\
 & 12 & 18 & 1 & 1742 & 370 & 4.71 & 1674 & 68 \\
 & 13 & 14 & 1 & 1549 & 370 & 4.19 & 1515 & 34 \\
 & 14 & 15 & 1 & 1622 & 370 & 4.38 & 1546 & 76 \\
 & 15 & 16 & 1 & 1647 & 370 & 4.45 & 1567 & 80 \\
 & 16 & 15 & 1 & 1610 & 370 & 4.35 & 1554 & 56 \\
 & 17 & 22 & 1 & 1639 & 370 & 4.43 & 1587 & 52 \\
 & 18 & 20 & 1 & 1592 & 370 & 4.3 & 1544 & 48 \\
 & 19 & 15 & 1 & 1621 & 370 & 4.38 & 1581 & 40 \\
 & 20 & 15 & 1 & 1645 & 370 & 4.45 & 1601 & 44 \\
 & 21 & 17 & 1 & 1584 & 370 & 4.28 & 1544 & 40 \\
 & 22 & 22 & 1 & 1614 & 370 & 4.36 & 1560 & 54 \\
 & 23 & 14 & 1 & 1578 & 370 & 4.26 & 1548 & 30 \\
 & 24 & 17 & 1 & 1670 & 360 & 4.64 & 1600 & 70 \\
 & 25 & 18 & 1 & 1545 & 360 & 4.29 & 1477 & 68 \\
 & 26 & 14 & 1 & 1538 & 360 & 4.27 & 1496 & 42 \\
 & 27 & 17 & 1 & 1598 & 360 & 4.44 & 1546 & 52 \\
 & 28 & 22 & 1 & 1571 & 360 & 4.36 & 1515 & 56 \\
 & 29 & 18 & 1 & 1514 & 353 & 4.29 & 1472 & 42 \\
\hline
\end{tabular}
\label{tab:planar-random-partition}
\end{table}

\subsection{Impact of Metis partitioning scheme.}
In this section, we consider the scenario when the user has additional information about the structure of input graph such as an efficient partitioning of the graph produced by Metis \cite{A-RK2006IPDPS}. Metis partitions reduce the edge-cut partitioning objective, i.e. the number of graph edges bridging different partitions, and thus increase the locality within the partitions. In other words, it helps reduce the amount of work for coordinating mutual exclusion between the clients. Consequently, the convergence of all execution modes is improved when compared to using the normal/sequential partitions. However, since the overhead of mutual exclusion is reduced, the benefit of aggressive stabilization mode EV-AS decreases (cf. Table \ref{tab:self-stabilization-vs-rollback-metis}).

\begin{table}[htbp]
\caption{Impact of Metis partitioning scheme.}
\begin{tabular}{|p{1.5cm}|p{2.8cm}|p{2cm}|p{2cm}|} 
\hline
& {\parbox{1.5cm}{Problem}} & \multicolumn{2}{c|}{\parbox{2cm}{Planar Graph Coloring}} \\ \hline
& Input graph & Planar & Planar \\ 
& Partition scheme & Sequential & Metis \\ 
\hline
\multirow{4}{*}{\parbox{1.5cm}{\centering{Conver-gence time (seconds)}}} & SEQ                   & 8,545 & 2,585 \\ 
& EVE-S                  & 6,173 & 2,389 \\ 
& EVE-AS                 & 2,590 & 2,154 \\ 
& Rollback               & 8,660 & 2,635 \\  
\hline
\multirow{4}{*}{\parbox{1.5cm}{\centering{Benefit}}}& EVE-S vs. SEQ       & 27.8\% & 7.6\% \\ 
& EVE-AS vs. SEQ     & $\times 3.3$ & $\times 1.2$ \\ 
& Rollback vs. SEQ   & -1.4\% & -1.9\% \\ 
\hline
\end{tabular}
\label{tab:self-stabilization-vs-rollback-metis}
\end{table}



\end{document}